\documentclass[doublecol]{epl2}
\usepackage{graphicx}
\usepackage[utf8,latin1]{inputenc}
\usepackage{dcolumn}
\usepackage{amsmath}
\usepackage{amssymb}
\usepackage[square,comma,numbers]{natbib}
\newcommand{\qm}[1]{``#1''}

\title{Analytical Rayleigh potential for the general relativistic Poynting-Robertson effect}

\author{Vittorio De Falco \and Emmanuele Battista}

\institute{Research Centre for Computational Physics and Data Processing, Faculty of Philosophy \& Science, Silesian University in Opava, Bezru\v{c}ovo n\'am.~13, CZ-746\,01 Opava, Czech Republic}
 
\abstract{We determine the analytic expression of the Rayleigh potential associated to the general relativistic Poynting-Robertson effect. This constitutes the first example of a physical dissipative system treated analytically in General Relativity. The employed approach and further implications are discussed.
}

\begin{document}

\maketitle

\section{Introduction}
Dissipative systems encompass a wide range of physical research areas, such as general relativity (GR), fluid dynamics, statistical mechanics, and quantum mechanics. Depending on the context, dissipation configures as a general mechanism through which a physical system may partially (or even completely) waste during time evolution its initial energy, entropy, entanglement, and so on \cite{Cantrijn1982,Ciaglia2018}. Dissipative contributions are fundamental expedients to make a model more realistic, even though the mathematical structure becomes more and more muddled.  

We consider the class of dissipative systems for which the equations of motion are known and exhibit both conservative and dissipative contributions. These can be schematically written as 
\begin{equation}\label{eq:PR_equat_motion_intr}
\boldsymbol{a}-\boldsymbol{F}_{\rm cons}=\boldsymbol{F}_{\rm diss},    
\end{equation}
where $\boldsymbol{a}$ is the acceleration, $\boldsymbol{F}_{\rm cons}$ and $\boldsymbol{F}_{\rm diss}$ the conservative and dissipative forces, respectively. The dynamics may be derived by the principle of least action through the Euler-Lagrange equations,
\begin{equation}
\frac{d}{dt}\left(\frac{\partial \mathcal{L}}{\partial \dot{q}^h}\right)-\frac{\partial \mathcal{L}}{\partial q^h}=Q_h^{\rm (nc)},\quad h=1,\dots,n,
\end{equation}
where $t$ is the evolution parameter (normally coincident with time), $\mathcal{L}(\boldsymbol{q},\boldsymbol{\dot{q}})$ the Lagrangian function containing the kinetic energy and the potential of all conservative forces, while $Q_h^{\rm (nc)}(\boldsymbol{q},\boldsymbol{\dot{q}})$
indicates all the forces not derivable from a potential. This leads to the renowned inverse problem in calculus of variation, which consists formally in the investigation of the necessary and sufficient conditions under which it is guaranteed existence and uniqueness of a Lagrangian description for a given second-order dynamics \cite{Santilli1978,Morandi1990,Do2016}. Matters complicate considerably when dissipation phenomena are included, since the notion of a \emph{well-posed problem} becomes a subtle task. \emph{Existence} of solution can be addressed by generalizing the Hirsch multiplier method (see Refs. \cite{Gitman2007b,Mestdag2011}). \emph{Uniqueness} 
leads to significant ambiguities affecting the implications of the Noether theorem and the quantization procedure (see Refs. 
\cite{deRitis1983,Gitman2007,Kochan2010}, and references therein). \emph{Stability of the solution with respect to initial data} requires several tough regularization schemes (see Ref. \cite{Kabanikhin2011}, for further details).

The approaches aimed at investigating problems involving dissipation constitute a huge domain. We classify such methods in three categories: \emph{quantitative} frameworks, represented by mathematical analytical techniques (see Refs. \cite{Evans2010,Brezis2010,Federer2014}), \emph{qualitative} patterns, relying on dynamical systems theory (see Refs. \cite{Sprott2011,Li2016}), and \emph{numerical} approaches (see Refs. \cite{Quarteroni2009,Quarteroni2015}). Although many of these classical techniques can be readily adapted to handle problems in metric theories of gravity, some of them turn out to be not fully incisive. Indeed, difficult problems are often entrusted through appropriate numerical means. Even though such tools configure as a precious resource, they sometimes discourage theoretical investigation. However, hybrid schemes, where dissipative phenomena are first analysed as dynamical systems through numerical simulations and afterwards specific analytic treatments are developed, should always be preferable. A suitable approach consists in encompassing dissipative phenomena by resorting to the Rayleigh dissipation potential \cite{Minguzzi2015}, which gives a direct insight in the mathematical and physical details of the problem under consideration. 

In this letter we have analytically determined the Rayleigh potential for the general relativistic Poynting-Robertson (PR) effect model. This represents a valuable achievement, since in the GR literature no attempts in such direction can be found. We start by describing the formal aspects of the approach we have pursued. After that, we exhibit the form of the Rayleigh potential function and discuss its peculiarities and further implications. 

\section{Method} We present the formal details of the pattern we have exploited to determine the analytic form of the Rayleigh potential of the general relativistic PR effect. The geometrical set up is based on a $4$-dimensional, real, topological Hausdorff, differential, pseudo Riemannian, asymptotically flat, and simply connected manifold $\mathcal{M}$. We define on $\mathcal{M}$: an atlas $\mathcal{A}$, a set of local coordinates $\boldsymbol{X}=(X^1,...,X^4)$, a metric tensor $\mathfrak{g}_{\alpha\beta}$, and a Lebesgue measure $\mathfrak{m}$. We denote with $T\mathcal{M}$ the tangent bundle of $\mathcal{M}$, whereas $T^*\mathcal{M}$ stands for the cotangent bundle over $\mathcal{M}$. $T\mathcal{M}$ is a simply connected domain, since it is the product of $\mathcal{M}$ and a topological vector space, which is always simply connected. The set of local coordinates defined on it are expressed by $(\boldsymbol{X},\boldsymbol{U})$ \cite{Neill1983,Nakahara2003,Lee2010}.

Let $\boldsymbol{\omega}:T\mathcal{M}\rightarrow T^*\mathcal{M}$ be a smooth\footnote{Smooth indicates here a function of class $\mathcal{C}^k(T\mathcal{M},\mathfrak{m})$ with $k\ge1$, i.e., $\mathfrak{m}$-continuous with the first $k$ derivatives $\mathfrak{m}$-continuous.}, differential semi-basic one-form. In local coordinates, we can write $\boldsymbol{\omega}(\boldsymbol{X},\boldsymbol{U})= F^\alpha(\boldsymbol{X},\boldsymbol{U})\ \boldsymbol{{\rm d}}X_\alpha$, where $F^\alpha$ are referred to as the components of $\boldsymbol{\omega}$. We assume that $\boldsymbol{\omega}$ is closed under the vertical exterior derivative $\boldsymbol{{\rm d^V}}$, i.e., $\boldsymbol{{\rm d^V}}\boldsymbol{\omega} = 0$ \cite{Godbillon1969,Abraham1978}. The local expression for this operator is given by
\begin{equation} \label{eq:vertical_derivative1}
\boldsymbol{{\rm d^V}}F= \frac{\partial F}{\partial U_\alpha} \boldsymbol{{\rm d}}X_\alpha, \quad \forall F\in \mathcal{C}^k(T\mathcal{M},\mathfrak{m}),\ k\ge1.
\end{equation}
Since the Poincar\'e lemma can be generalised also to the case of vertical differentiation \cite{Martinez1992}, and bearing in mind our hypotheses regarding the topological property of $T\mathcal{M}$, the closure condition guarantees that $\boldsymbol{\omega}$ is exact. Therefore, it can be expressed as the vertical exterior derivative of a 0-form $V(\boldsymbol{X},\boldsymbol{U})\in \mathcal{C}^k(T\mathcal{M},\mathfrak{m})$ (with $k\ge1$), refereed to as the primitive or the potential function, through $-\boldsymbol{{\rm d^V}} V=\boldsymbol{\omega}$. Now, we consider the following \emph{energy function}:
\begin{equation} \label{eq:constraint}
\mathbb{E}=\rho(\boldsymbol{X},\boldsymbol{U}),
\end{equation}
where $\rho$ is a smooth real valued function of the coordinates $(\boldsymbol{X},\boldsymbol{U})$, and $\mathbb{E}$ represents the energy dissipated by the system. Equation (\ref{eq:constraint}) permits to introduce in the components of the differential semi-basic one-form $\boldsymbol{\omega}(\boldsymbol{X},\boldsymbol{U})$ the explicit dependence on the energy $\mathbb{E}$, i.e.,
\begin{equation} \label{eq:constraint_2}
F^\alpha=F^\alpha(\mathbb{E},\boldsymbol{X},\boldsymbol{U}).
\end{equation}
It should be stressed that in the Eq. (\ref{eq:constraint_2}) we have not altered the number of variables, i.e., $F^\alpha(\mathbb{E},\boldsymbol{X},\boldsymbol{U})$ is still function of the $8$ local coordinates on $T\mathcal{M}$. As we will see in the next section, we have obtained such relation by simply substituting all the occurrences of $\rho$ with $\mathbb{E}$. 

Equations (\ref{eq:constraint}) and (\ref{eq:constraint_2}) represent the key aspects of the strategy we have employed, since they turn out to be useful in simplifying the demanding calculations involved in the search for the $V$ primitive. However, we stress that Eq. (\ref{eq:constraint}) must be interpreted neither as a constraint nor as the first integral of a conservative system, aimed at lowering the number of dynamical equations to be solved. However, Eq. (\ref{eq:constraint}) leads to a powerful result, because it allows to reduce substantially the coordinates occurring in the subsequent computations, passing from $4$ initial parameters, represented by $\boldsymbol{U}$, to one only, i.e., the energy $\mathbb{E}$. Another fundamental step consists in passing from the velocity to the energy derivative operator through the usual chain rule, i.e., 
\begin{equation} \label{eq:trader}
\frac{\partial\ (\ \cdot\ )}{\partial U_\alpha}=\frac{\partial \rho}{\partial U_\alpha}\ \frac{\partial\ (\ \cdot \ )}{\partial \mathbb{E}},
\end{equation}
so that the $V$ function satisfies the condition
\begin{equation} \label{eq:primitive_original}
F^\alpha=-\frac{\partial \rho}{\partial U_\alpha}\frac{\partial V}{\partial \mathbb{E}},
\end{equation}
where the usual definition of primitive function has been employed. Such differential equation for $V$ contains the $\partial \rho/\partial U_\alpha$ factor, which might represent an obstacle for the integration process. To get rid of this term, we can consider the scalar product of both members of Eq. (\ref{eq:primitive_original}) by an appropriate function $B(\boldsymbol{X},\boldsymbol{U})_\alpha$, which permits to obtain a more manageable integral equation for $V$:
\begin{equation} \label{eq:pot_E}
V=\int \left(\frac{-F^\alpha B_\alpha}{\frac{\partial \rho}{\partial U_\alpha}B_\alpha} \right){\rm d}\mathbb{E}+f(\boldsymbol{X},\boldsymbol{U}),
\end{equation}
where $f$ is constant with respect to $\mathbb{E}$, i.e., $\partial f/\partial \mathbb{E}=0$ and, according to the discussion following Eq. (\ref{eq:constraint_2}), $V$ is still a function of the local coordinates $(\boldsymbol{X},\boldsymbol{U})$.

The $f$ function occurring in Eq. (\ref{eq:pot_E}) fulfils a crucial role in our method, since it embodies the whole information of the $V$ potential in terms of the $\boldsymbol{U}$ coordinates. The occurrence of such unknown function should not be considered as something unexpected, since it represents the logical consequence of having employed Eq. (\ref{eq:constraint}), which allows to enhance the role of the physical variable $\mathbb{E}$ with respect to the others. In other words, our ignorance about the analytic form of $V$ as a function of the local coordinates $(\boldsymbol{X},\boldsymbol{U})$ is \qm{buried} in $f(\boldsymbol{X},\boldsymbol{U})$, which can thus be interpreted as a gauge function of $V$ with respect to $\mathbb{E}$. However, the  $f$ function can be determined by applying the usual iterative process for the computation of the primitive of an exact differential one-form. 

\section{General relativistic PR effect} 
We now outline how the pattern described in the previous section has allowed us to derive the Rayleigh potential of PR effect. 
This phenomenon occurs in high-energy astrophysics when radiation processes are considered. Indeed, the radiation field produced by the emitting source, beside exerting an outward radial force opposite to the gravitational attraction, produces also a drag force, which is due to the process of absorption and reemission of the radiation undergone by the affected body \cite{Poynting1903,Robertson1937}. The drag force removes very efficiently angular momentum and energy from the body, thus altering its motion.

The PR model deals with the dynamics of a test particle orbiting with a timelike velocity $\boldsymbol{U}$ a rotating compact object, under the influence of the gravitational field, described by the Kerr metric, the radiation pressure, and the PR radiation drag force \cite{Bini2009,Bini2011,Defalco2018,Defalco20183D}. The latter can be interpreted as a dissipative force. The test particle equations of motion read as (cf. Eq. (\ref{eq:PR_equat_motion_intr}))
\begin{equation}
a(\boldsymbol{X},\boldsymbol{U})^\alpha=F_{\rm (rad)}(\boldsymbol{X},\boldsymbol{U})^\alpha,
\end{equation}
where $a(\boldsymbol{X},\boldsymbol{U})^\alpha$ is the test particle acceleration and $F_{\rm (rad)}(\boldsymbol{X},\boldsymbol{U})^\alpha$ is the radiation force per unit mass, including radiation pressure and PR effect. This force is modelled as a pure electromagnetic field, where photons move along null geodesics of the Kerr spacetime,  i.e., 
\begin{equation} \label{eq:radforce1}
F_{\rm (rad)}(\boldsymbol{X},\boldsymbol{U})^\alpha=-\tilde{\sigma}\Phi^2\left(k^\alpha k_\nu U^\nu+U^\alpha U_\beta k^\beta k_\nu U^\nu\right),
\end{equation}
where $\tilde{\sigma}=\sigma/m$ with $\sigma$ the Thomson scattering cross section describing the radiation field-test particle interaction and $m$ the test particle mass, $\Phi$ represents a parameter related to the radiation field intensity, $k^\alpha$ denotes the photon 4-momentum. 

The geometrical setup of the PR effect fully respects the hypothesis described in the previous section. Therefore, it is meaningful to search for the Rayleigh potential function $V(\boldsymbol{X},\boldsymbol{U})$. Indeed, in the PR model, the manifold $\mathcal{M}$ denotes the spacetime outside the compact object (i.e, $\mathcal{M}$ is simply connected) and $\mathfrak{g}_{\alpha\beta}$ is represented by the Kerr metric. Furthermore, the Lebesgue measure coincides with the standard measure of length, area, or volume in Kerr spacetime. The radiation force components (\ref{eq:radforce1}) are the components of the differential semi-basic one-form $\boldsymbol{\omega}(\boldsymbol{X},\boldsymbol{U})=F_{\rm (rad)}(\boldsymbol{X},\boldsymbol{U})^\alpha \boldsymbol{{\rm d}}X_\alpha$, defined over the simply connected domain $T\mathcal{M}$. The energy function (\ref{eq:constraint}) is given by the dissipated energy $\mathbb{E}=-k_\beta U^\beta$ of the test particle. 

Due to the non-linear dependence of the radiation force on the test particle velocity field, the semi-basic one-form turns out to be not exact \cite{Defalco2018}. However, the PR phenomenon exhibits the peculiar propriety according to which $\boldsymbol{\omega}(\boldsymbol{X},\boldsymbol{U})$ becomes exact through the introduction of the integrating factor\footnote{In $\mu$ we have corrected a little error occurred in Ref. \cite{Defalco2018}.}
$\mu =\left(E_{\rm p}/\mathbb{E}\right)^2$, where the photon energy $E_{\rm p}$ is a constant term determined in the classical limit. By exploiting Eqs. (\ref{eq:constraint_2})--(\ref{eq:pot_E}), we have derived the analytical form assumed by the Rayleigh potential of the PR effect. The final result reads as 
\begin{equation} \label{eq: Rayleigh_potential_final}
V=\tilde{\sigma}\Phi^2\left[\ln\left(\frac{\mathbb{E}}{E_{\rm p}}\right)+\frac{1}{2}\left(U_\alpha U^\alpha+1\right)\right],
\end{equation}
where the constant term $\lg{E_{\rm p}}-1/2$ has been determined in the classical limit. Equation (\ref{eq: Rayleigh_potential_final}) is consistent with the classical description \cite{Poynting1903,Robertson1937}. Therefore, PR effect configures as the first dissipative model in the context of Einstein theory for which an analytical Rayleigh potential has been found. 

\section{Results and discussions}
We have presented for the first time an analytical expression of the Rayleigh potential for a dissipative system in GR. 
The potential function (\ref{eq: Rayleigh_potential_final}) shows a peculiar behaviour due to the presence of the logarithmic term $\log(\mathbb{E}/E_p)$, which turns out to be well suited to the description of radiation absorption processes in high-energy astrophysics.
Indeed, as the radiation source emits photons towards the test particle, the faster the test particle moves, the less the photons are absorbed. This is reflected in the fact that when the test particle moves at the speed of light $\mathbb{E}\to0$ and hence the potential term $\log(\mathbb{E}/E_p)\to-\infty$. On the contrary, if the test particle is at rest, the absorption is maximum, because $\mathbb{E}=E_p$ and $\log(\mathbb{E}/E_p)=0$. This implies that the absorption term occurring in the Rayleigh potential is always negative, because $\mathbb{E}\le E_p$.

In our approach the introduction of an integrating factor is a fundamental expedient \cite{Defalco2018}. This term has a deep meaning, since it is connected with the energy of the system. Moreover, it reduces to a constant in the classical limit, indicating that it gives a significant dynamical contribution only at relativistic level. The function $f(\boldsymbol{X},\boldsymbol{U})$ appearing in Eq. (\ref{eq:pot_E}) has been easily determined in the case of the PR effect. However, its introduction may become fundamental in handling different classes of dissipative systems. Indeed, in general the computation of $f(\boldsymbol{X},\boldsymbol{U})$ might require progressively lengthier calculations, especially when the dimension of $\mathcal{M}$ increases. Nevertheless, this may be exactly the point where our method comes into its own, since even if such calculations turn out to be unfeasible, we can always achieve the \emph{analytic} expression of $V$ in terms of the variable $\mathbb{E}$, which represents the actual physical quantity enabling a thorough description of the dynamics, especially when dissipation phenomena occur. In other words, instead of trying to express $V$ in terms of local coordinates by means of some complex approach, we could set forth a pattern to find the analytic form of $V$ as a function of the variable having a precise physical meaning, at the cost of shifting our ignorance to a function whose unknown nature does not prevent a complete description of the dynamics. 

In order to fully appreciate this strategy, we note that it shows remarkable similarities with the approach originally pursued by D'Alembert and Lagrange in order to analyse (free or constrained) mechanical systems. Indeed, its crucial aspect consists on the proof that a limited set of essential physical variables (i.e., the Lagrangian coordinates) are sufficient to determine the dynamics. The method we are proposing works analogously except that, and this aspect might reveal its significant breakthrough, it deals with dissipative systems. Indeed, Eq. (\ref{eq:constraint}) permits to recognise the energy $\mathbb{E}$ as the only fundamental physical parameter necessary to express analytically the $V$ primitive and hence describe the dynamics.

The above considerations may represent the source of possible generalizations of the method enlightened in this letter.
Indeed, the great simplicity by which it makes possible to obtain an analytic form of the potential function in a  dissipative system in GR surely represents a good result. Moreover, being defined within a geometrical setup, it can be widely framed both in theoretical and applied physics, as well as in pure mathematical analysis. For instance, we note that our scheme applies not only to differential semi-basic one-form, but also to differential one-forms and higher-order differential one-forms (defined on the $m$-th-order tangent bundle $T^m\mathcal{M}$ of $\mathcal{M}$, $m$ being an integer such that $m\ge1$ \cite{Lee2010}), provided that all definitions stated here are appropriately recasted. Another remarkable characteristic is that this approach is \emph{metric-independent}, i.e., it is not altered by the background geometry, but acts exclusively on the functional form of the involved equations. This ensures its validity in GR and, in a broader sense, in any metric theory of gravity.       

\section{Conclusions}
\label{sec:conclusions}

In this letter, we have analytically determined the Rayleigh dissipation function for the general relativistic PR draf force. Its final form is reported in Eq. (\ref{eq: Rayleigh_potential_final}). At the best of our knowledge, such a result has never been established in the literature regarding inverse problems involving dissipation phenomena in the calculus of variation in GR. The method we have adopted is mainly based on the introduction of the energy function (\ref{eq:constraint}) and the $f(\boldsymbol{X},\boldsymbol{U})$ function (see Eq. (\ref{eq:pot_E})), which has made the underlying calculations more feasible.

The analytic form of the Rayleigh potential is a valuable tool for various reasons, since it allows, e.g., to grasp essential details of radiative processes within the realm of potential theory. In particular, it would make possible a straight connection between theoretical model and observations. This topic is under active consideration at the present time. 
Indeed, the functional form of the Rayleigh potential can be used to investigate neatly the dynamical features of a radiation process and vice versa observational data can be exploited in order to infer the proprieties of the Rayleigh potential. This paves the way to several interesting future applications both in theoretical and applied physics.

There are various aspects of the strategy we have outlined that could make it general and hence valid for the analysis of dissipative systems different from the PR model. Indeed, it has been worked out by exploiting the powerful and elegant machinery of differential geometry and besides it turns out to be metric independent. Moreover, the technique of the integrating factor could be further enforced in metric theories of gravity to enlarge the class of integrable dissipative systems. In particular, the investigation and the characterisation of this functional class may constitute a new research field to be pursued in future works. Finally, the introduction in Eq. (\ref{eq:pot_E}) of the $f(\boldsymbol{X},\boldsymbol{U})$ function may represent a suited compromise in those dynamical systems where the research of the potential requires demanding calculations.   

In conclusion, the approach we have adopted in this letter, which we may call ``energy formalism'', offers interesting hints pointing towards its possible generalizations. This feature may open up a new research window in the field of the inverse problems involving dissipation in metric theories of gravity.



\acknowledgments
The authors are grateful to Professor Luigi Stella, Doctor Giampiero Esposito, Professor Antonio Romano, Professor Giuseppe Marmo, and Professor Tom Mestdag for the stimulating discussions and the useful suggestions aimed at improving the scientific impact of this letter. The authors thank the Silesian University in Opava and the International Space Science Institute in Bern for hospitality and support. The authors are grateful to Gruppo Nazionale di Fisica Matematica of Istituto Nazionale di Alta Matematica for support. 

\bibliographystyle{eplbib}
\bibliography{references}

\end{document}